\documentclass[copyright]{eptcs}
\providecommand{\event}{PLACES 2013}

\usepackage{graphicx}
\usepackage{listings}
\usepackage{caption}
\usepackage{subcaption}
\usepackage{wrapfig}

\newcommand{\Csharp}{C$^\sharp$}
\newcommand{\Fsharp}{F$^\sharp$}

\title{Minimising virtual machine support for concurrency}
\author{Simon Dobson \qquad Alan Dearle \qquad Barry Porter
  \institute{School of Computer Science, University of St Andrews UK}
  \email{simon.dobson@st-andrews.ac.uk}}
\def\titlerunning{Minimising VM support for concurrency}
\def\authorrunning{Dobson, Dearle, and Porter}

\begin{document}
\maketitle

\begin{abstract}
  Co-operative and pre-emptive scheduling are usually considered to be
  complementary models of threading. In the case of virtual machines,
  we show that they can be unified using a single concept, the
  \emph{bounded execution} of a thread of control, essentially
  providing a first-class representation of a computation as it is
  reduced. Furthermore this technique can be used to surface the
  thread scheduler of a language into the language itself, allowing
  programs to provide their own schedulers without any additional
  support in the virtual machine, and allowing the same virtual
  machine to support different thread models simultaneously and
  without re-compilation.
\end{abstract}

\section{Introduction}
\label{sec:introduction}

Multiple threads of control can make programs easier to write, by
allowing logically concurrent activities to be coded independently,
regardless of the availability of true concurrency. These benefits
accrue equally to all levels of the software stack: applications,
operating systems and virtual machines. In multicore environments
there can also be performance benefits, and there are almost always
extra complexities due to the need to control simultaneous or
interleaved access to resources -- but improved expressiveness alone
makes multiple threads attractive even for single-cored systems

Virtual machines (VMs) can further improve expressiveness, by providing a
base level of abstraction targeted to the needs of the higher-level
languages being written and hiding the underlying
complexities of different machine architectures. A typical virtual
machine embodies a particular model of threading and hard-codes a
particular model of thread scheduling, prioritisation and so on. This
choice will almost always be sub-optimal for some class of
applications, especially in the case of resource-constrained embedded
systems and sensor networks where it may be desirable for applications
to exert close control over all aspects of the system's
operation. Equally, we want to keep the VM well-defined and not
shuffle important features into platform-dependent libraries.

In this paper we observe that it is possible to construct a virtual
machine that makes no \textit{a priori} choices about thread
scheduling and concurrency control, yet without relegating these vital
functions to external libraries. Put another way, we allow the
\textit{same} VM to support \emph{different} concurrency models. We do
this by simplifying the VM's support for threading to a single
function offering \textit{bounded execution} of the virtual
instruction stream. This approach allows us to surface all other
aspects of concurrency out of the VM and into the language.

\section{Virtual machines and concurrency}
\label{sec:related-work}

Several modern programming language implementations adopt a virtual
machine approach. The most notable are Java (the Java Virtual Machine
or JVM); \Csharp{} and \Fsharp{} (the Common Language
Infrastructure~\cite{ECMA-CLI}); GNU and Squeak Smalltalk; Lua; and
Python. These VMs all adopt the \emph{bytecode} style, in which the VM
defines the instruction set of a processor that is ``ideal'' in some
sense for the language being defined. This focus allows the language
and VM designers to collaborate to define an instruction set that
exactly matches the needs of the language, and so minimise space, time
and compilation overheads, and the trade-offs between them. Some VMs
are general enough to be targeted by several languages: by design in
the case of CLI, or through ingenuity in the case of the JVM targeted
by Java and Scala. It also allows different implementations of the
same instruction set, targeted at different classes of machine (for
example for Java standard~\cite{InsideJVM}, bare-metal~\cite{SquawkVM}
and just-in-time~\cite{EllulRunTimeCompilation} VMs, and Smalltalk's
highly portable and long-lived object engine and associated
VM~\cite{SmalltalkBackToTheFuture}).

\begin{wrapfigure}{r}{.5\textwidth}
  \lstset{language=C}
  \begin{lstlisting}
    /* VM */
    void bytecode_inner_interpreter() {
      while(TRUE) {
        bytecode = *ip++;
        opcode = unpack_opcode(bytecode);
        switch(opcode) {
          case OP_NOOP:
            break;
          ...
          case OP_JUMP:
            ip += unpack_offset(bytecode);
            break;
          ...
        }
      }
    }
  \end{lstlisting}

  \caption{Interpreting bytecode}
  \label{fig:bytecode-inner-interpreter}
\end{wrapfigure}
The core of a virtual machine is an \emph{inner interpreter} that
identifies and interprets virtual instructions. For a bytecode VM the
inner interpreter uses a virtual instruction pointer (IP) to read an
instruction and select an appropriate behaviour to execute. The
bytecode may be ``packed'' to include (for example) a small integer
literal or a jump offset to allow common instructons to take up less
space, and so may require some decoding to extract the \emph{opcode}
specifying the instruction's implementation (a \emph{primitive})
before execution
(figure~\ref{fig:bytecode-inner-interpreter})\footnote{In the code
  fragments included in this paper, we use C syntax both for the code
  \textit{implementing} the VM and the target language code
  \emph{implemented using} the VM. We differentiate between the two
  using a leading comment.}.

In a multithreaded architecture, each
thread must give up the processor after some time (referred to as its
\emph{time quantum}) to allow another thread to execute. In a
\emph{co-operative} (or coroutine) scheduler the programmer embeds
explicit instructions that yield control at programmer-selected
execution points. This slightly complicates programming and means that
poorly-written code may not yield often enough (or at all) to avoid
\emph{starving} other threads of processor time, but has the advantage
of requiring little or no concurrency control on data structures,
since a thread may manipulate shared data without fear of interference
from another thread. (In single-core systems, at least: multi-core
requires slightly more care, for example separate control structures
for each core.) By contrast, a \emph{pre-emptive} scheduler forcibly
interrupts the executing thread, suspending it and allowing another
thread to be selected. This gives more power to the scheduler and
prevents starvation, but requires concurrency control over all shared
data.

Typically each thread maintains its own stack space and instruction
pointer cache. Changing threads (a \emph{context switch}) saves the
VM's stack and instruction pointers and replaces them with those of
the newly-scheduled thread. Implementing co-operative scheduling on a
VM is conceptually straightforward, with a \texttt{yield()} primitive
invoking a context switch. Pre-emptive scheduling often leads VM
designers to use OS-level threads, swapping between several different
VM-level instruction streams. Context switches also need to be
triggered when a thread blocks on some event, such as a semaphore or a
file read. This reduces the control the VM can offer to the language
level, and often makes its behaviour platform-specific. (Java, for
example, has primitives to block on object semaphores, but none for
thread creation or scheduling~\cite{InsideJVM}.)

\subsection{Approaches to VM concurrency}
\label{sec:vm-concurrency}

Despite being such an important feature of a programming language,
many VMs -- perhaps surprisingly -- leave their concurrency model
under-specified. The JVM, for example, includes bytecodes for
concurrency \emph{control}, manipulating the locks associated with
objects, but none for concurrency \emph{creation}. This has the
advantage of leaving a particular JVM implementation free to make use
of native libraries to implement threads, but simultaneously removes
the issue from the VM's design. Most JVMs run concurrent instances of
themselves in different OS-defined threads, accepting the overhead in
terms of concurrency control within the VM that this
imposes. Smalltalk's VM similarly omits instructions for concurrency
control.

Some implementations of Scheme have adopted the ``engines''
construct~\cite{HayesFriedmanEngines}, which re-defines the lambda
binder to maintain a timer (essentially counting the number of
function calls made) that is then used to interrupt a thread when a
certain number of reductions have occurred. It is possible to
implement engines inside Scheme, on top of a standard implementation
(whether using a VM or otherwise), by making use of continuation
capture~\cite{DybvigHiebEngines}. Care needs to be taken to ensure
that such an implementation is loaded first so as to capture all
lambdas, and there is also the risk that such an approach will
conflict with other uses of continuation capturing.

\section{Bounded concurrency control}
\label{sec:bounded-control}

We assume a single core and therefore no ``true'' concurrency. Similar
techniques can be used in multicore environments.

The core problem in context switching is to take control away from a
running language-level thread (either voluntarily or forcibly) and give
it to another. We propose to accomplish this by changing the inner
interpreter so that -- instead of running an \emph{unbounded} loop
over a single virtual instruction stream which must be interrupted to
regain control -- it runs a \emph{bounded} loop to context-switch into
a given thread and execute a certain maximum number of virtual
instructions before returning
(figure~\ref{fig:simple-bounded-inner-interpreter}). Here
\texttt{activate()} context-switches the VM's instruction pointer and
other registers to those of a given thread. If we assume that
all primitives are non-blocking, then the such a bounded inner
interpreter will always return to its caller in a finite
time. Non-blocking primitives are a strong assumption, but
multithreading actually simplifies the creation of blocking structures
on a non-blocking substrate by allowing condition checking to be made
independent of the main program flow (exactly as an operating system does).

\begin{wrapfigure}{r}{.5\textwidth}
  \lstset{language=C}
  \begin{lstlisting}
    /* VM */
    void bounded( bound, thread ) {
      oldthread = activate(thread);
      n = bound;
      while(n--) {
        bytecode = *ip++;
        opcode = unpack_opcode(bytecode);
        switch(opcode) {
          ...
        }
      }
      activate(oldthread);
    }
  \end{lstlisting}

  \caption{Bounded interpretation}
  \label{fig:simple-bounded-inner-interpreter}
\end{wrapfigure}
We could simply use this construction to avoid the need for
interruption and to re-factor the scheduler into another primitive
that uses the bounded inner interpreter. However, the construction
facilitates a more interesting approach whereby we remove the
\emph{entire} scheduling and concurrency control regime from the VM
and surface it to the language level.

Having bounded the inner interpreter, we may now treat it as a virtual
instruction in its own right (since it is non-blocking). This means
that we may use it in defining new behaviours, and specifically we may
use it in thread scheduling and concurrency control.  The significance
of this change is two-fold. Firstly, it blurs the distinction between
co-operative and pre-emptive thread scheduling. Suppose we provide
language-level threads, each of which is represented as a VM-level
thread. At the VM level, thread scheduling is essentially
co-operative: the bounded inner interpreter runs the thread for a time
quantum specified in terms of virtual instructions and performs a
voluntary context switch. At the language level, however, threads are
pre-empted arbitrarily (from their perspective), since they have no
control over when the underlying VM will switch them out.  Secondly,
bounded execution means that thread scheduling can itself be provided
by a thread, rather than as a primitive. The scheduling thread chooses
a worker thread, boundedly executes it for its time quantum, receives
control back and selects another (or the same) thread for
execution. Thread scheduling therefore need not be considered as a
primitive function of the VM, and may instead happen at language
level: one language-level thread can use bounded execution to run
another for a given period, without losing overall control of the
program's execution.

\begin{wrapfigure}{r}{.5\textwidth}
  \lstset{language=C}
  \begin{lstlisting}
    /* Target language */
    while(TRUE) {
        task = dequeue(runqueue);
        bounded(quantum, task);
        enqueue(task, runqueue);
    }
  \end{lstlisting}

  \caption{Language-level scheduling}
  \label{fig:simple-round-robin}
\end{wrapfigure}

\subsection{Thread creation and scheduling}

A thread is created by allocating memory for its stacks and essential
registers -- tasks that can be performed without primitive support --
before scheduling the thread by adding it to the scheduler's run
queue. We might encode the simplest round-robin scheduler as shown in
figure~\ref{fig:simple-round-robin}. The point is that this is
\emph{program} code and not VM code: it need not be primitive, and so
may be redefined independently of the VM.

Within this style more complex schedulers are clearly possible. A
scheduler might maintain multiple run queues of differing priorities
and select the next thread from the highest-priority queue having
runnable threads. The \texttt{quantum} parameter
determines the latency of context switches in terms of virtual
instructions: one might reduce this number to regain control into the
scheduler more frequently, and consult a timer to determine whether to
perform a context switch, leading to language-level threads that
effectively have time quanta specified in wallclock times (at some
granularity) rather than in virtual instructions.

\subsection{Program-level concurrent objects}

A small modification of the bounded inner interpreter allows us to
migrate semphores and control of other program-level concurrent
objects to the language level alongside the scheduler.

If we ignore the possibility that the thread may be pre-empted,
implementing semaphores at language level is straightforward. A
semaphore consists of a counter and a thread queue. The wait (P)
function decrements the counter and, if it is less than zero, enqueues
the thread onto the thread queue and de-schedules it as far as the
thread scheduler is concerned. The signal (V) operation increments the
counter and, if it remains less than zero, dequeues a thread from the
semaphore's thread queue and adds it to the scheduler's run
queue. None of these functions require explicit VM support.

Dealing with pre-emption requires that we change the bounded inner
interpreter in three ways. Firstly, we add a state flag to each thread
which by default is \texttt{RUNNABLE} indicating that the thread may
continue to execute. Secondly, we introduce two other states:
\texttt{BLOCKED} for a thread that is blocked on a thread queue and so
cannot be scheduled; and \texttt{PRIORITISED} for a thread that should
not be pre-empted. Setting a thread's state to \texttt{PRIORITISED}
forces the bounded inner interpreter to keep executing virtual
instructions in this thread, even if it comes to the end of its
allocated quantum. Finally, we return the thread state from the
bounded inner interpreter. This new scheme is shown in
figure~\ref{fig:bounded-inner-interpreter}, and is VM-level code:
\texttt{set\_thread\_state()} is another primitive that
sets the running thread's state. We modify the scheduler (at language
level) so that it runs prioritised and, after receiving control back
from \texttt{bounded()}, it only enqueues the thread back onto the run
queue if it is \texttt{RUNNABLE} (figure~\ref{fig:round-robin}). (Note
that the scheduler is target language, non-primitive code.)

\begin{figure}[chtp]
  \centering
  \begin{subfigure}[b]{.45\textwidth}
    \lstset{language=C}
    \begin{lstlisting}
    /* VM */
    int bounded( bound, thread ) {
      oldthread = activate(thread);
      n = bound;
      set_thread_state(RUNNABLE);
      while((state = thread_state()),
            (state == PRIORITISED) ||
            ((state != BLOCKED) && (n--))) {
          opcode = unpack_opcode(bytecode);
          switch(opcode) {
              ...
          }
      }
      activate(oldthread);
      return state;
    }
    \end{lstlisting}

    \caption{Inner interpreter}
    \label{fig:bounded-inner-interpreter}
  \end{subfigure}
  \\[1cm]
  \begin{subfigure}[b]{.45\textwidth}
    \lstset{language=C}
    \begin{lstlisting}
    /* Target language */
    set_thread_state(PRIORITISED);
    while(TRUE) {
        task = dequeue(runqueue);
        state = bounded(quantum, task);
        if(state == RUNNABLE)
            enqueue(task, runqueue);
    }
    \end{lstlisting}

    \caption{Round-robin scheduler}
    \label{fig:round-robin}
  \end{subfigure}

  \caption{Language-level concurrency control}
\end{figure}

We can now write a \texttt{wait()} primitive (for example) at language
level rather than as a VM primitive. We first set the thread's state
to \texttt{PRIORITISED}. The thread can then manipulate the
semaphore's counter and thread queue safely using the full features of
the language, since it will not be pre-empted. At the end of the
definition we set the thread's state to \texttt{RUNNABLE} if we pass
the semaphore or \texttt{BLOCKED} if the thread has been enqueued on the
semaphore's thread queue. A \texttt{RUNNABLE} thread will keep running
or, if it has reached its quantum, will be de-scheduled and enqueued
on the run queue; a \texttt{BLOCKED} thread will not be enqueued. The
complementary implmentation of \texttt{signal()} will prioritise the
thread, dequeue a blocked thread (if any) from the sempahore, enqueue
it on the run queue, and then make itself \texttt{RUNNABLE} again to
restore normal scheduling.

What this shows is that the bounded inner interpreter with
simple atomic thread state-setting offers sufficient VM-level support to
allow concurrency primitives to be lifted to language level, and
so allow a program to take complete control of its own concurrency
control regime. These operations then have access to the full scope of
the language: they are not restricted to the functions available
primitively within the VM.

Clearly there is scope for errors in this scheme if program code is
allowed to arbitrarily make itself \texttt{PRIORITISED}, which
essentially turns the pre-emptive scheme into a co-operative scheme
again. However, this is a problem for the language level that may be
addressed using permissions, encapsulation or whatever mechanisms (if
any) the designer chooses: it is not an issue for the VM, which can
support any scheme chosen.

\subsection{Considerations of VM design}

At the design level, the VM does not need to have visibility of either
the scheduling policy or the mechanisms (queues, semaphores, etc) used
to implement it: all such considerations can be raised into the target
language, to be re-implemented as required. This is in contrast to the
more standard approaches to VM implementation
(section~\ref{sec:vm-concurrency}) in which the same structures are
\emph{submerged} into the run-time system that underlies the VM. This
places far more control in the hands of the language designer and
implementer.

This approach to concurrency is orthogonal to any other mechanisms
provided within the VM, with the single proviso that all primitives be
non-blocking so as to be properly schedulable. The scheduling
\emph{mechanism} is applied ``below'' the target language rather than
``within'' it, even while the scheduling \emph{policy} is provided
within the target: it does not re-use a language-level feature
(such as continuation capture) that might also be used in other ways
that interfere with the scheduling mechanism.

It is interesting to note the way in which bounded concurrency
highlights the closeness of the two concepts of thread and
continuation, with a continuation capturing the future of a
computation compared to a thread capturing the on-going reduction of
that computation.

\section{Conclusion}
\label{sec:conclusion}

We have briefly presented an approach to opening-up the concurrency
mechanisms in a virtual machine, allowing the VM to provide minimal
support (two primitive operations) and building the rest of the
concurrency regime at the language level. We have shown that
this supports a number of different approaches to concurrency,
including allowing the definition of new thread schedulers within a
language so that they can be changed and specialised at run-time. One
may also choose between traditional and speculative concurrency,
blocking and non-blocking data structures and the like, entirely
on top of the VM and therefore completely portably. 

This scheme allows us to further enrich the program-level handling of
concurrency on top of a minimal virtual machine. It is possible, for
example, to unify the treatment of threads and delimited
continuations, making these powerful features available on-demand with
little or no VM support -- and therefore no overhead where they are
not required. This is potentially of great significance for sensor networks and
other systems with severely limited resources, and we are currently
exploring what place such advanced language features have in such
environments.

\bibliographystyle{eptcs}
\bibliography{places-2013-eptcs}

\end{document}